\documentclass[aps,prl,twocolumn,superscriptaddress,showpacs,preprintnumbers,amsmath,amssymb]{revtex4-1}

\usepackage[left]{lineno}
\usepackage{graphicx} 
\usepackage{subfigure}	
\usepackage{dcolumn}  
\usepackage{eqnarray}
\graphicspath{{ps}}

\widetext
\begin{document}

\title{ \quad\\[1.0cm] Search for $B^0 \to X(3872) \gamma$}

\noaffiliation
\affiliation{University of the Basque Country UPV/EHU, 48080 Bilbao}
\affiliation{Beihang University, Beijing 100191}
\affiliation{Brookhaven National Laboratory, Upton, New York 11973}
\affiliation{Budker Institute of Nuclear Physics SB RAS, Novosibirsk 630090}
\affiliation{Faculty of Mathematics and Physics, Charles University, 121 16 Prague}
\affiliation{Chonnam National University, Kwangju 660-701}
\affiliation{University of Cincinnati, Cincinnati, Ohio 45221}
\affiliation{Deutsches Elektronen--Synchrotron, 22607 Hamburg}
\affiliation{Duke University, Durham, North Carolina 27708}
\affiliation{University of Florida, Gainesville, Florida 32611}
\affiliation{Key Laboratory of Nuclear Physics and Ion-beam Application (MOE) and Institute of Modern Physics, Fudan University, Shanghai 200443}
\affiliation{Justus-Liebig-Universit\"at Gie\ss{}en, 35392 Gie\ss{}en}
\affiliation{Gifu University, Gifu 501-1193}
\affiliation{SOKENDAI (The Graduate University for Advanced Studies), Hayama 240-0193}
\affiliation{Gyeongsang National University, Chinju 660-701}
\affiliation{Hanyang University, Seoul 133-791}
\affiliation{University of Hawaii, Honolulu, Hawaii 96822}
\affiliation{High Energy Accelerator Research Organization (KEK), Tsukuba 305-0801}
\affiliation{J-PARC Branch, KEK Theory Center, High Energy Accelerator Research Organization (KEK), Tsukuba 305-0801}
\affiliation{Forschungszentrum J\"{u}lich, 52425 J\"{u}lich}
\affiliation{IKERBASQUE, Basque Foundation for Science, 48013 Bilbao}
\affiliation{Indian Institute of Science Education and Research Mohali, SAS Nagar, 140306}
\affiliation{Indian Institute of Technology Bhubaneswar, Satya Nagar 751007}
\affiliation{Indian Institute of Technology Guwahati, Assam 781039}
\affiliation{Indian Institute of Technology Hyderabad, Telangana 502285}
\affiliation{Indian Institute of Technology Madras, Chennai 600036}
\affiliation{Indiana University, Bloomington, Indiana 47408}
\affiliation{Institute of High Energy Physics, Chinese Academy of Sciences, Beijing 100049}
\affiliation{Institute of High Energy Physics, Vienna 1050}
\affiliation{INFN - Sezione di Napoli, 80126 Napoli}
\affiliation{INFN - Sezione di Torino, 10125 Torino}
\affiliation{Advanced Science Research Center, Japan Atomic Energy Agency, Naka 319-1195}
\affiliation{J. Stefan Institute, 1000 Ljubljana}
\affiliation{Institut f\"ur Experimentelle Teilchenphysik, Karlsruher Institut f\"ur Technologie, 76131 Karlsruhe}
\affiliation{Kennesaw State University, Kennesaw, Georgia 30144}
\affiliation{King Abdulaziz City for Science and Technology, Riyadh 11442}
\affiliation{Department of Physics, Faculty of Science, King Abdulaziz University, Jeddah 21589}
\affiliation{Kitasato University, Sagamihara 252-0373}
\affiliation{Korea Institute of Science and Technology Information, Daejeon 305-806}
\affiliation{Korea University, Seoul 136-713}
\affiliation{Kyoto University, Kyoto 606-8502}
\affiliation{Kyungpook National University, Daegu 702-701}
\affiliation{LAL, Univ. Paris-Sud, CNRS/IN2P3, Universit\'{e} Paris-Saclay, Orsay}
\affiliation{\'Ecole Polytechnique F\'ed\'erale de Lausanne (EPFL), Lausanne 1015}
\affiliation{P.N. Lebedev Physical Institute of the Russian Academy of Sciences, Moscow 119991}
\affiliation{Liaoning Normal University, Dalian 116029}
\affiliation{Faculty of Mathematics and Physics, University of Ljubljana, 1000 Ljubljana}
\affiliation{Ludwig Maximilians University, 80539 Munich}
\affiliation{Luther College, Decorah, Iowa 52101}
\affiliation{University of Malaya, 50603 Kuala Lumpur}
\affiliation{University of Maribor, 2000 Maribor}
\affiliation{Max-Planck-Institut f\"ur Physik, 80805 M\"unchen}
\affiliation{School of Physics, University of Melbourne, Victoria 3010}
\affiliation{University of Mississippi, University, Mississippi 38677}
\affiliation{University of Miyazaki, Miyazaki 889-2192}
\affiliation{Moscow Physical Engineering Institute, Moscow 115409}
\affiliation{Moscow Institute of Physics and Technology, Moscow Region 141700}
\affiliation{Graduate School of Science, Nagoya University, Nagoya 464-8602}
\affiliation{Kobayashi-Maskawa Institute, Nagoya University, Nagoya 464-8602}
\affiliation{Universit\`{a} di Napoli Federico II, 80055 Napoli}
\affiliation{Nara Women's University, Nara 630-8506}
\affiliation{National Central University, Chung-li 32054}
\affiliation{National United University, Miao Li 36003}
\affiliation{Department of Physics, National Taiwan University, Taipei 10617}
\affiliation{H. Niewodniczanski Institute of Nuclear Physics, Krakow 31-342}
\affiliation{Nippon Dental University, Niigata 951-8580}
\affiliation{Niigata University, Niigata 950-2181}
\affiliation{Novosibirsk State University, Novosibirsk 630090}
\affiliation{Osaka City University, Osaka 558-8585}
\affiliation{Pacific Northwest National Laboratory, Richland, Washington 99352}
\affiliation{Panjab University, Chandigarh 160014}
\affiliation{Peking University, Beijing 100871}
\affiliation{University of Pittsburgh, Pittsburgh, Pennsylvania 15260}
\affiliation{Punjab Agricultural University, Ludhiana 141004}
\affiliation{Theoretical Research Division, Nishina Center, RIKEN, Saitama 351-0198}
\affiliation{University of Science and Technology of China, Hefei 230026}
\affiliation{Seoul National University, Seoul 151-742}
\affiliation{Showa Pharmaceutical University, Tokyo 194-8543}
\affiliation{Soongsil University, Seoul 156-743}
\affiliation{University of South Carolina, Columbia, South Carolina 29208}
\affiliation{Stefan Meyer Institute for Subatomic Physics, Vienna 1090}
\affiliation{Sungkyunkwan University, Suwon 440-746}
\affiliation{School of Physics, University of Sydney, New South Wales 2006}
\affiliation{Department of Physics, Faculty of Science, University of Tabuk, Tabuk 71451}
\affiliation{Tata Institute of Fundamental Research, Mumbai 400005}
\affiliation{Department of Physics, Technische Universit\"at M\"unchen, 85748 Garching}
\affiliation{Toho University, Funabashi 274-8510}
\affiliation{Department of Physics, Tohoku University, Sendai 980-8578}
\affiliation{Earthquake Research Institute, University of Tokyo, Tokyo 113-0032}
\affiliation{Department of Physics, University of Tokyo, Tokyo 113-0033}
\affiliation{Tokyo Institute of Technology, Tokyo 152-8550}
\affiliation{Tokyo Metropolitan University, Tokyo 192-0397}
\affiliation{Virginia Polytechnic Institute and State University, Blacksburg, Virginia 24061}
\affiliation{Wayne State University, Detroit, Michigan 48202}
\affiliation{Yamagata University, Yamagata 990-8560}
\affiliation{Yonsei University, Seoul 120-749}
  \author{P.-C.~Chou}\affiliation{Department of Physics, National Taiwan University, Taipei 10617} 
  \author{P.~Chang}\affiliation{Department of Physics, National Taiwan University, Taipei 10617} 
  \author{I.~Adachi}\affiliation{High Energy Accelerator Research Organization (KEK), Tsukuba 305-0801}\affiliation{SOKENDAI (The Graduate University for Advanced Studies), Hayama 240-0193} 
  \author{H.~Aihara}\affiliation{Department of Physics, University of Tokyo, Tokyo 113-0033} 
  \author{S.~Al~Said}\affiliation{Department of Physics, Faculty of Science, University of Tabuk, Tabuk 71451}\affiliation{Department of Physics, Faculty of Science, King Abdulaziz University, Jeddah 21589} 
  \author{D.~M.~Asner}\affiliation{Brookhaven National Laboratory, Upton, New York 11973} 
  \author{H.~Atmacan}\affiliation{University of South Carolina, Columbia, South Carolina 29208} 
  \author{V.~Aulchenko}\affiliation{Budker Institute of Nuclear Physics SB RAS, Novosibirsk 630090}\affiliation{Novosibirsk State University, Novosibirsk 630090} 
  \author{T.~Aushev}\affiliation{Moscow Institute of Physics and Technology, Moscow Region 141700} 
  \author{R.~Ayad}\affiliation{Department of Physics, Faculty of Science, University of Tabuk, Tabuk 71451} 
% ...
\collaboration{The Belle Collaboration}

\noaffiliation

\begin{abstract}
We report the results of a search for the decay $B^0 \to X(3872)(\to J/\psi \pi^+ \pi^-) \gamma$. The analysis is performed on a data sample corresponding to an integrated luminosity of $711\,{\rm fb}^{-1}$  and containing $772 \times 10^6 B\bar{B}$ pairs, collected with the Belle detector at the KEKB asymmetric-energy $e^+ e^-$ collider running at the $\Upsilon(4S)$ resonance energy.
We find no evidence for a signal and place an upper limit of $\mathcal{B}(B^0 \to X(3872)\gamma)\times \mathcal{B}(X(3872) \to J/\psi \pi^+ \pi^-) < 5.1 \times 10^{-7}$ at 90\% confidence level.
\end{abstract}


\maketitle

\tighten

{\renewcommand{\thefootnote}{\fnsymbol{footnote}}}
\setcounter{footnote}{0}

\begin{figure}[htb]
\includegraphics[width=0.27\textwidth]{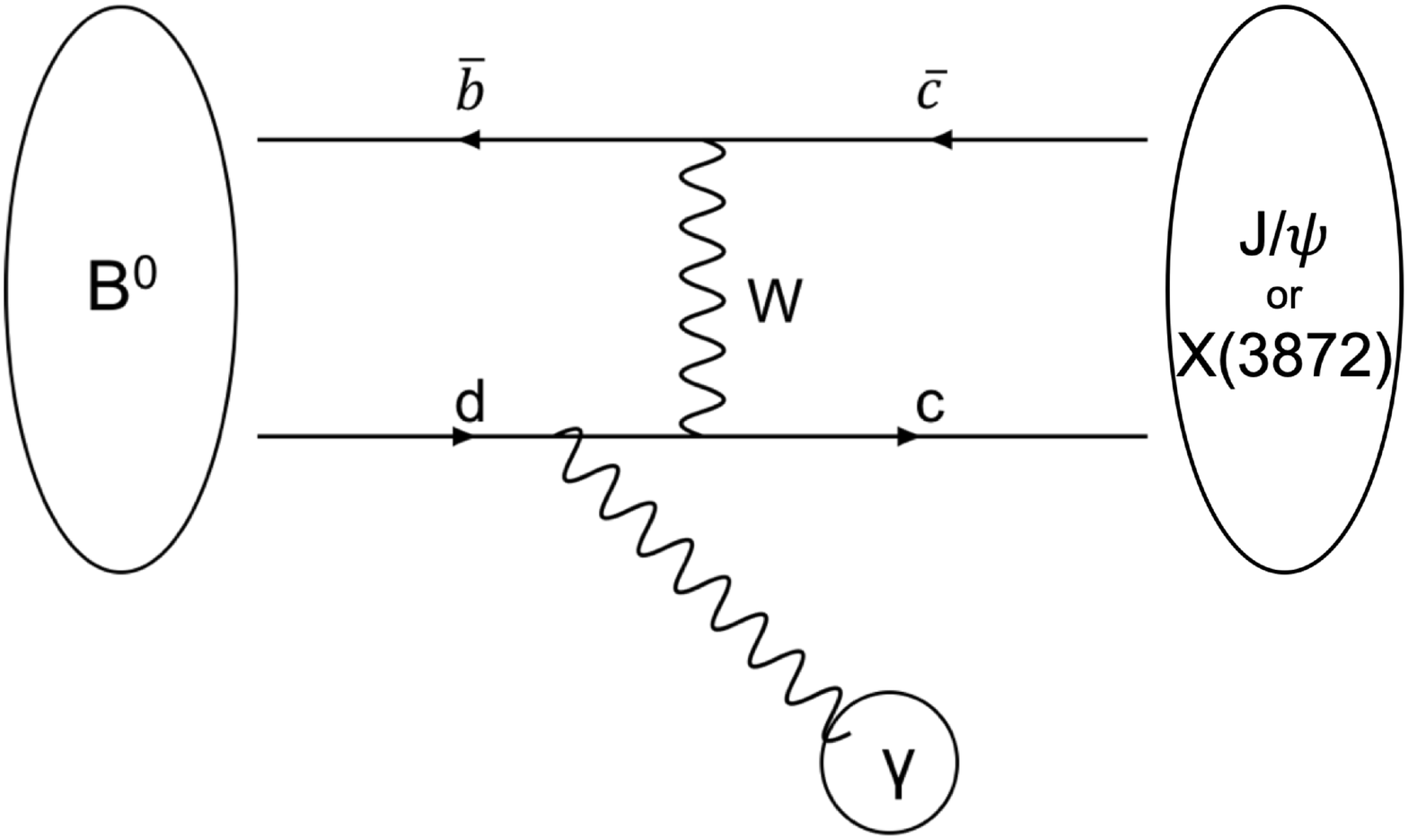}
\caption{A Feynman diagram of $B^0 \to c \bar{c} \gamma$.}
\label{fig:feyn}
\end{figure}

Rare decays of $B$ mesons are sensitive probes to study possible new physics beyond the Standard Model (SM). In the SM, the decay $B^0 \to c\bar{c} \gamma$ proceeds dominantly through an exchange of a $W$ boson and the radiation of a photon from the $d$ quark of the $B$ meson (Fig.~\ref{fig:feyn}). Many theoretical predictions of branching fractions depend on the factorization approach of QCD interactions in the decay dynamics. In the case of $B^0 \to J/\psi \gamma$, the branching fraction has been predicted to be $7.65 \times 10^{-9}$ using QCD factorization~\cite{ref:Eur2004} and $4.5 \times 10^{-7}$ when using a perturbative QCD (pQCD) approach~\cite{ref:PRD2006}. 
Possible new physics enhancements of the branching fractions may be due to right-handed currents~\cite{ref:Eur2004} or nonspectator intrinsic charm in the $B^0$ meson~\cite{ref:PRD2002}. Currently, the upper limit for $B^0 \to J/\psi \gamma$ is $1.5 \times 10^{-6}$ at 90\% confidence level~\cite{ref:PDG18}. 

The exotic $X(3872)$ state, first observed by the Belle experiment in 2003~\cite{ref:PRL2003}, is now one of the most well-studied charmoniumlike exotic states.
Aside from pure charmonium, it may also be a $D^0 \bar{D}^{*0}$ molecule~\cite{ref:Swanson2004}, a tetraquark state~\cite{ref:Maiani2005}, or a mixture of a molecule and a charmonium~\cite{ref:Suzuki}.
Since $X(3872)$ may, unlike the $J/\psi$, contain components other than pure charmonium, the branching fraction of $B^0 \to X(3872) \gamma$ should be smaller than that of $B^0 \to J/\psi \gamma $ which proceeds through the $b\to c \bar{c}$ $d$ process. No former search for this decay has been published yet.

Our measurement is based on a data sample corresponding to an integrated luminosity of $711\,{\rm fb}^{-1}$  that contains $772 \times 10^6 B\bar{B}$ pairs, 
collected with the Belle detector at the KEKB asymmetric-energy $e^+e^-$ (3.5 on 8~GeV) collider~\cite{KEKB} running at the $\Upsilon(4S)$ resonance. 
The Belle detector is a large-solid-angle magnetic
spectrometer that consists of a silicon vertex detector (SVD),
a 50-layer central drift chamber (CDC), an array of
aerogel threshold Cherenkov counters (ACC),  
a barrellike arrangement of time-of-flight
scintillation counters (TOF), and an electromagnetic calorimeter
comprised of CsI(Tl) crystals (ECL), all located inside 
a superconducting solenoid coil that provides a 1.5~T
magnetic field.  An iron flux-return yoke located outside
the solenoid is instrumented to detect and identify $K_L^0$ mesons and muons (KLM).  The detector
is described in detail elsewhere~\cite{Belle}.

Two inner detector configurations are used in this analysis. A beam pipe of radius 2.0 cm
and a 3-layer SVD were used for the first data set of $140\,{\rm fb}^{-1}$, while a beam pipe of 1.5 cm radius and a 4-layer SVD silicon detector were used to record the remaining data set of $571\,{\rm fb}^{-1}$~\cite{svd2}. All the Monte Carlo (MC) samples in this analysis are generated by the \textsc{EvtGen} package~\cite{EvtGen} and the response of the Belle detector is simulated by the \textsc{Geant3} package~\cite{Geant}.  QED final-state radiation (FSR) is simulated using the \textsc{photos} package~\cite{PHOTOS}. These samples are used to optimize the selection criteria and determine the signal detection efficiency.
Simulation assumes that the $X(3872)$ decays to
$J/\psi\pi^+\pi^-$ entirely via $J/\psi\rho^0$.
The $J/\psi$ then decays to two channels: $\mu^+ \mu^-$ or $e^+ e^-$. Charge-conjugate modes are implied throughout this paper. We generate one million events for each channel. In any given event of signal MC, only one of the two $B$ mesons will decay via the signal mode while the other $B$ meson decays generically. A signal MC event is considered a correctly reconstructed one if it matches the particle type and the momentum difference between the signal MC and the reconstructed tracks is less than 0.05 GeV$/c$.

Selection criteria for the final-state charged particles in $B^0 \to X(3872) \gamma$ are based on information obtained from the tracking systems (CDC and SVD) and the hadron identification systems (CDC, ACC, and TOF). 
Charged final-state particles are identified using information provided by the CDC, the TOF, the ACC, the ECL, and the KLM. 
The pion candidates are identified using information from the ACC (number of photoelectrons), the CDC ($dE/dx$), and the TOF. 
The muon candidates are identified using track penetration depth and hit information in the KLM. 
The electron candidates are identified using the transverse shape and size of the showers in the ECL, the CDC ($dE/dx$), the ACC, the ratio of ECL energy to the CDC track momentum, and the position matching between the CDC track and the ECL cluster.
These pieces of information are combined to form a likelihood $\mathcal{L}$ for charged particle identification.

We require $\pi^\pm$ candidates to satisfy $\mathcal{L}_{\pi/K} = \mathcal{L}_\pi/(\mathcal{L}_\pi+\mathcal{L}_K)> 0.6$, while rejecting highly electronlike ($\mathcal{L}_e > $ 0.95) or muonlike ($\mathcal{L}_\mu > $ 0.95) tracks.
For muon tracks, we require the particle identification likelihood $\mathcal{L}_\mu > 0.9$. 
We define electron tracks as those with the particle identification likelihood $\mathcal{L}_e > 0.9$.
Charged tracks are required to originate from the nominal interaction point, which can avoid poorly measured tracks or tracks which do not come from $B$ decays. For charged pion tracks, we require the impact parameters in the radial direction ($dr$) and in the beam direction ($dz$) to satisfy $dr < 2.0$ cm and $|dz| < 5.0$ cm, respectively. For lepton tracks, we require $|dr| < 0.2$ cm and $|dz| < 2.0$ cm.

We reconstruct $J/\psi$ candidates in the $\ell^+\ell^-$ decay channel ($\ell \in \{e,\mu\}$) and include bremsstrahlung photons that are within 50 mrad of the $e^+$ or $e^-$ tracks. The invariant mass window used to select $J/\psi$ candidates in the $\mu^+\mu^-$($e^+e^-$) channel is 3.03 (2.95) GeV/$c^2 \leq M_{\mu\mu}(M_{ee}) \leq $ 3.13 GeV/$c^2$. 
These intervals are asymmetric in order to include parts of the radiative tails. 
A lower mass requirement for the $e^+e^-$ channel is used because electron tracks are more sensitive to energy loss due to bremsstrahlung. 
The lower bound corresponds to 1.7 and 1.4 standard deviations for dimuon and dielectron channel, respectively; 
the upper bound corresponds to 2.5 and 2.1 standard deviation for dimuon and dielectron channel, respectively.
We also require $\chi^{2}_{\ell\ell}/\text{n.d.f.} < 20$, where $\chi^{2}_{\ell\ell}/\text{n.d.f.}$ is the $\chi^{2}$ per degree of freedom of the $J/\psi \to \ell^+\ell^-$ vertex fit.
The $J/\psi$ candidate is then combined with a $\pi^+\pi^-$ pair to reconstruct an $X(3872)$ candidate. The invariant mass windows used to select $X(3872)$ candidates are 3.7 GeV/$c^2 < M_{\mu\mu\pi\pi} < $ 3.95 GeV/$c^2$ and 3.5 GeV/$c^2 < M_{ee\pi\pi} < $ 3.95 GeV/$c^2$. We require the di\-pion invariant mass to satisfy $M_{\pi\pi} > M_{\ell\ell\pi\pi} - M_{\ell\ell} - 150$ MeV/$c^2$. This selection was introduced in an earlier analysis~\cite{ref:BN1177} to reduce combinatorial backgrounds from misidentified $\gamma$ conversions, which correspond to $M_{\pi\pi} > 625$ MeV/$c^2$ for the $X(3872)$. 
After the $M_{\pi\pi}$ selection is applied, about 15.9\% (for dimuon channel) and 15.8\% (for dielectron channel) of the true signal is removed, while about 43.0\% (for dimuon channel) and 42.7\% (for dielectron channel) of the combinatorial background in the signal region is rejected. 
The $\chi^2$/n.d.f. of the $\rho^0$ vertex fit is constrained within $\chi^{2}_{\pi\pi}/\text{n.d.f.} < 80$. Selections on $\Delta M = M_{\ell\ell\pi\pi} - M_{\ell\ell}$ can also be employed to reduce combinatorial backgrounds. We require 0.755 GeV/$c^2 < M_{\mu\mu\pi\pi} - M_{\mu\mu} < $ 0.795 GeV/$c^2$ and 0.745 GeV/$c^2 < M_{ee\pi\pi} - M_{ee}< $ 0.805 GeV/$c^2$. After the $\Delta M$ selection is applied, about 0.73\%  (for di\-muon channel) and 0.43\% (for di\-electron channel) of the true signal is removed, while about 80.8\% (for di\-muon channel) and 75.2\% (for di\-electron channel) of the combinatorial background in the signal region is rejected. 
The value of $\chi^2$/n.d.f.\ of the $X(3872)$ vertex fit is required to be within $\chi^{2}_{\ell\ell\pi\pi}/\text{n.d.f.} < 100$.

A high-energy photon produces an electromagnetic shower in the ECL, and it is detected as an isolated energy cluster which is not associate with charged particles. 
The energy of the photon candidate coming from the $B^0$ is required to be larger than 0.6 GeV in the center-of-mass (CM) frame. We also reject the photon candidate if the ratio of the energies deposited in arrays of $3\times3$ and $5\times5$ calorimeter cells ($E_9/E_{25}$) is less than 0.87. 
To reduce background from the decay $\pi^0 \to \gamma \gamma$, a $\pi^0$ veto is applied with $\mathcal{L}_{\pi^0}<0.3$, where $\mathcal{L}_{\pi^0}$ is a $\pi^0$ likelihood~\cite{ref:koppenberg}. 
The $B$ meson candidate is then reconstructed by combining the $X(3872)$ candidate and the high-energy photon candidate.
 $B$ meson candidates are identified with kinematic variables calculated in the CM frame (and denoted with an asterisk *).
The energy difference is calculated as $\Delta E=E^*_\text{recon} - E^*_\text{beam}$, where $E^*_\text{recon}$ and $E^*_\text{beam}$ are the reconstructed $B$ meson energy and beam energy. 
We use a modified beam-energy-constrained mass $M_\text{bc}$ defined as
\begin{equation}
M_\text{bc} = \sqrt{\Big(\frac{E^*_\text{beam}}{c^2}\Big)^2 - \Big( \frac{\vec{P}^*_{X}}{c}+\frac{\vec{P}^*_\gamma}{|\vec{P}^*_\gamma| c^2}(E^*_\text{beam}-E^*_\text{X})\Big)^2},
\end{equation}
where $\vec{P}^*_{X}$ and $E^*_{X}$ are the reconstructed momentum and energy of the $X(3872)$ candidate, and $\vec{P}^*_\gamma$ is the reconstructed momentum of the photon candidate.
The use of this modified definition reduces the linear correlation between the $M_\text{bc}$ and $\Delta E$ from 0.400 (0.332) to $-0.013$ (0.114) for the di\-muon (di\-electron) channel, as estimated with MC signal events. This also improves $M_\text{bc}$ resolution. 
The selection region is defined by $M_\text{bc} >$ 5.2 GeV/$c^2$ and $-0.5$ GeV $< \Delta E <$ 0.2 GeV. The signal region is defined by $M_\text{bc} >$ 5.27 GeV/$c^2$ and $-0.15$ GeV $< \Delta E <$ 0.1 GeV.

There are two main types of background events: the generic $B \overline{B}$ spherical events and the jetlike $q\overline{q}$ continuum events. 
The dominant background in the selection region is from the $B \to J/\psi X$ inclusive decays.
Other types of $B \overline{B}$ and continuum backgrounds also contribute.
Since the signal and background shapes are different, we use a multivariate analyzer based on the neural network package named \textsc{NeuroBayes}~\cite{NB} to distinguish the signal and background. We train the neural network using the signal MC and $B \to J/\psi X$ MC samples, with the following 33 input variables: 
(1) 25 modified Fox-Wolfram moments treating the information of particles involved in the signal $B$ candidate separately from those in the rest of the event~\cite{SFW},
(2) the cosine of the angle between the $B$ candidate and the beam axis, 
(3) the angle between the thrust axis of the decay particles of the $B$ candidate and that of the remaining particles in the event, 
(4) the event sphericity~\cite{spher}, 
(5) the missing mass, momentum and energy in the event,
(6) the sum of the transverse energy of the event, and 
(7) the flavor tagging information~\cite{TaggingNIM}.
Variables (1)--(6) are calculated in the $\Upsilon(4S)$ rest frame. 
\textsc{NeuroBayes} returns an output in the range $-1$ to +1, where values closer to +1 are signallike and values closer to $-1$ are backgroundlike. The applied selection on the \textsc{NeuroBayes} output is determined by optimizing a figure of merit (FOM)  defined as
\begin{equation}
{\rm FOM} = \frac{\text{efficiency}}{0.5n+\sqrt{N_\text{bkg}}},
\end{equation}
where $N_\text{bkg}$ is the number of background events and efficiency is obtained from the signal MC. In this equation, $n$ is the number of standard deviations corresponding to one-sided Gaussian tests, and $n=1.28$ corresponds to 90\% confidence level~\cite{ref:fom}. The optimized selection and its related systematic uncertainty are channel dependent.

If multiple candidates are found in an event after background suppression, we select the candidate which has the smallest $|\Delta M - 775$ MeV$/c^2|$. Before applying the selection, the multiplicity per event is 1.080 for the di\-muon channel and 1.116 for the di\-electron channel in signal MC samples. After the selection is applied, about 1.8\%  (for the di\-muon channel) and 2.0\% (for the di\-electron channel) of the true signal is removed, and about 43.1\% (for the di\-muon channel) and 47.0\% (for the di\-electron channel) of the combinatorial background in the signal region is rejected. With all of the selections applied, the di\-muon signal MC sample comprises 92\% correctly-reconstructed signal $B$ events (``true" signal) and 8\% self-crossfeed (SCF) events (not correctly reconstructed ones), and the di\-electron sample comprises 89\% true signal and 11\% SCF events.

The branching fraction is calculated as
\begin{equation}
\mathcal{B} = \frac{N_\text{sig}}{\epsilon \times \eta \times N_{B\bar{B}}},
\end{equation}
where $N_\text{sig}$, $N_{B\bar{B}}$, $\epsilon$ and $\eta$ are the number of signal, the number of $B\bar{B}$ pairs ($ = 772\times 10^6$), the signal reconstruction efficiency, and an efficiency calibration factor, respectively.
We assume that the charged and neutral $B\bar{B}$ pairs are equally produced at the $\Upsilon(4S)$. 

The calibration factor  
$\eta = \eta_\text{NB} \times \eta_{\pi \text{ID}} \times \eta_{\ell \text{ID}} \times \eta_{\pi^0 \text{veto}}\times\eta_\text{box}$ 
is a correction factor to the Monte Carlo that has been determined using real data and following methods:
$\eta_\text{NB}$ concerns the background suppression using 
\textsc{NeuroBayes} and is obtained using the $B^0 \to J/\psi (\to \ell^+ \ell^-) K_S^0$
control sample with treating $K_S^0$ as $\gamma$. We also check 
using another control sample $B^0 \to \psi(2S) (\to J/\psi \pi^+ \pi^-) K_S^0$,
which as a topology more similar to the signal, to verify the result. The two methods are in good agreement.
$\eta_{\pi \text{ID}}$ concerns the charged pion identification with the requirement on $\mathcal{L}_{\pi}$, and is determined using a $D^{*+} \to D^0 (\to K^- \pi^+) \pi^+$ control sample, 
$\eta_{\ell \text{ID}}$ concerns the lepton identification with the requirement on $\mathcal{L}_{\mu}$ or $\mathcal{L}_{e}$, and is determined by using a $e^+ e^- \to e^+ e^- \ell^+ \ell^-$ control sample with $e^+ e^-$ undetected, and
$\eta_{\pi^0 \text{veto}}$ concerns the $\pi^0$ veto with the requirement on $\mathcal{L}_{\pi^0}$, and is determined using a $B^0 \to D^- (\to K^+ \pi^- \pi^-) \pi^+$ control sample.
$\eta_\text{box}$ concerns the fraction of the signal yield in the signal region to that in the selection region after all selection is applied, and is determined by using a $B^0 \to K^0_S \pi^+ \pi^- \gamma$ control sample. 
The values of the calibration factors and the reconstructed efficiency for the signal with all the selection criteria applied are listed in Table \ref{table:calib}.

\begin{table}[htb]
\caption{Calibration factors ($\eta$) and reconstructed efficiency ($\epsilon$) for the signal with all the selection criteria applied.}
\label{table:calib}
\begin{tabular}
{@{\hspace{0.5cm}}l@{\hspace{0.5cm}}  @{\hspace{0.5cm}}c@{\hspace{0.5cm}}  @{\hspace{0.5cm}}c@{\hspace{0.5cm}}}
\hline \hline
Channel & Di\-muon  & Di\-electron\\
\hline
$\eta_\text{NB}$ & $0.98 \pm 0.02$ & $0.99 \pm 0.03$ \\
$\eta_{\pi \text{ID}}$ & $0.99 \pm 0.01$ & $0.99 \pm 0.01$ \\
$\eta_{\ell \text{ID}}$  & $0.96 \pm 0.02$ & $0.98 \pm 0.02$ \\
$\eta_{\pi^0 \text{veto}}$  & $0.98 \pm 0.01$ & $0.98 \pm 0.01$\\
$\eta_\text{box}$  & $0.95\pm0.03$ & $0.95\pm0.03$\\
\hline
$\eta$ & $0.86 \pm 0.06$ & $0.89 \pm 0.06$ \\
\hline
$\epsilon$ & $(16.8\pm 0.01)\%$ & $(14.5\pm 0.01)\%$ \\
\hline \hline
\end{tabular}
\end{table}

Sources of various systematic uncertainties on the branching fraction calculation are shown in Table \ref{table:sys}. 
The uncertainty due to the total number of $B\bar{B}$ pairs is 1.4\%.
The uncertainty due to the charged-track reconstruction efficiency is estimated to be 0.35\% per track 
by using partially reconstructed $D^{*+} \to D^0(\pi^+\pi^-K^0_S)\pi^+$ decay samples.
The uncertainty due to the subdecay $J/\psi \to \ell^+ \ell^-$ branching fraction is based on the world average value~\cite{ref:PDG18}.
The uncertainty due to the photon detection efficiency in the barrel region ($33^{\circ} < \theta_\gamma < 128^{\circ}$, where $\theta_\gamma$ is the polar angle of the photon) is studied using a radiative Bhabha sample,
and $B^0 \to K^{*0} \gamma$ elsewhere~\cite{ref:photoneff}.
The uncertainty due to the $X(3872) \to J/\psi \rho^0$ generation model is studied by comparing the signal MC samples generated with helicity distributions $\cos\theta$ (which is taken for the central value of efficiency), $\sin^2\theta$, and $1+\cos^2\theta$.

\begin{table}[htb]
\caption{Summary of systematic uncertainties on the measured branching fraction.}
\label{table:sys}
\begin{tabular}
{@{\hspace{0.5cm}}l@{\hspace{0.5cm}}  @{\hspace{0.5cm}}c@{\hspace{0.5cm}}  @{\hspace{0.5cm}}c@{\hspace{0.5cm}}}
\hline \hline
Source & Di\-muon  & Di\-electron\\
\hline
$N_{B\bar{B}}$ & 1.4\% & 1.4\% \\
Tracking (4 tracks) & 1.4\% & 1.4\% \\
$\mathcal{B}(J/\psi \to \ell^+ \ell^-)$ & 0.6\% &  0.5\% \\
$\gamma$ detection   & 3.1\% & 3.1\% \\
MC gen. model    & 1.1\% & 1.9\% \\
$\pi^\pm$ identification  & 1.3\% & 1.3\% \\
$\ell^\pm$ identification   & 2.1\% & 1.8\% \\
Bkg. suppression   & 2.3\% & 2.5\% \\
$\pi^0$ veto   & 0.8\% & 0.8\% \\
Signal region fraction & 3.5\% & 3.5\% \\
\hline
Total & 6.2\% & 6.4\% \\
\hline \hline
\end{tabular}
\end{table}

\begin{figure}[htb]
\includegraphics[width=0.48\textwidth]{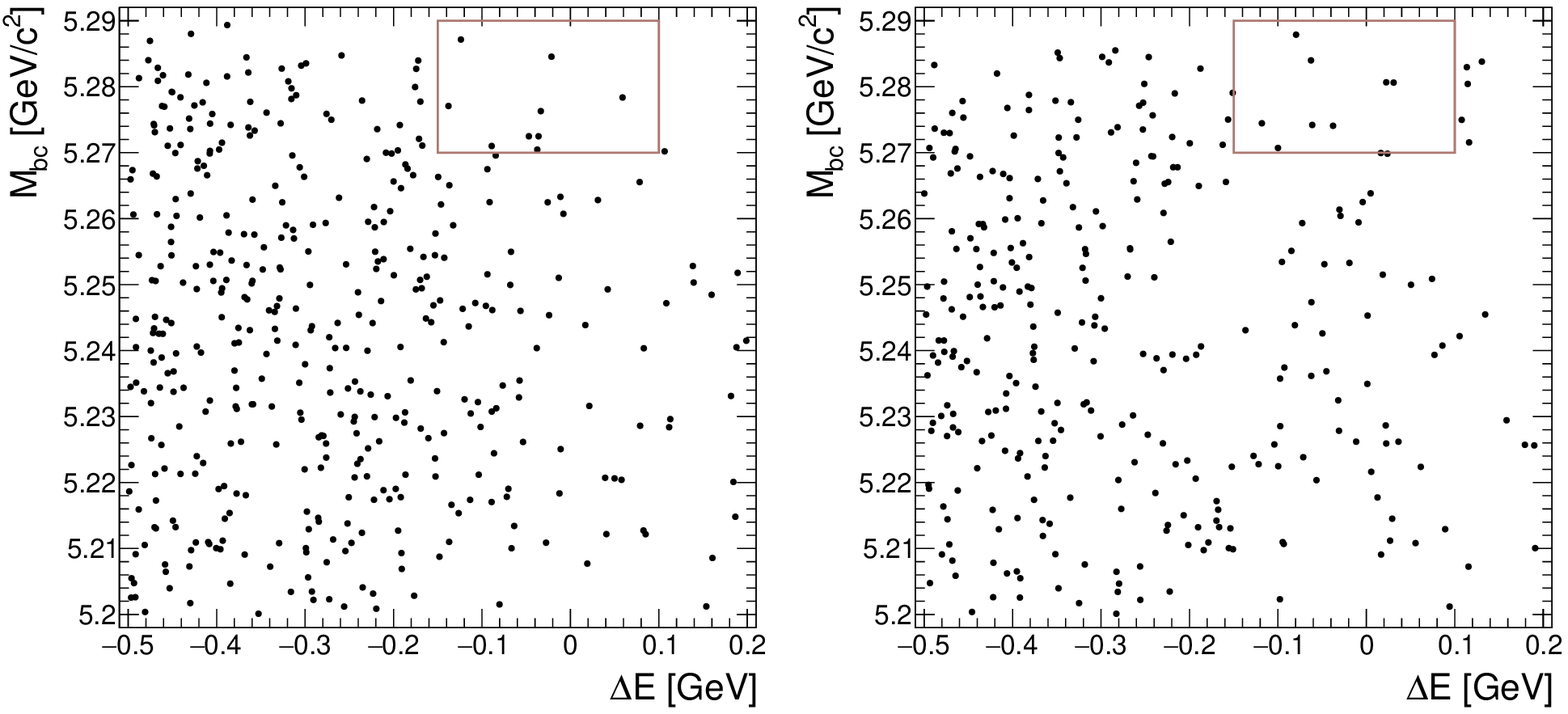}
\caption{Two dimensional ($M_\text{bc}$,$\Delta E$) distributions of the selected $B^0 \to X(3872) \gamma$ candidates in the di\-electron (left) and di\-muon (right) channels. The signal regions are shown as rectangles.}
\label{fig:scat}
\end{figure}

\begin{figure}[htb]
\subfigure[Di\-muon channel.]{
\includegraphics[width=0.5\textwidth]{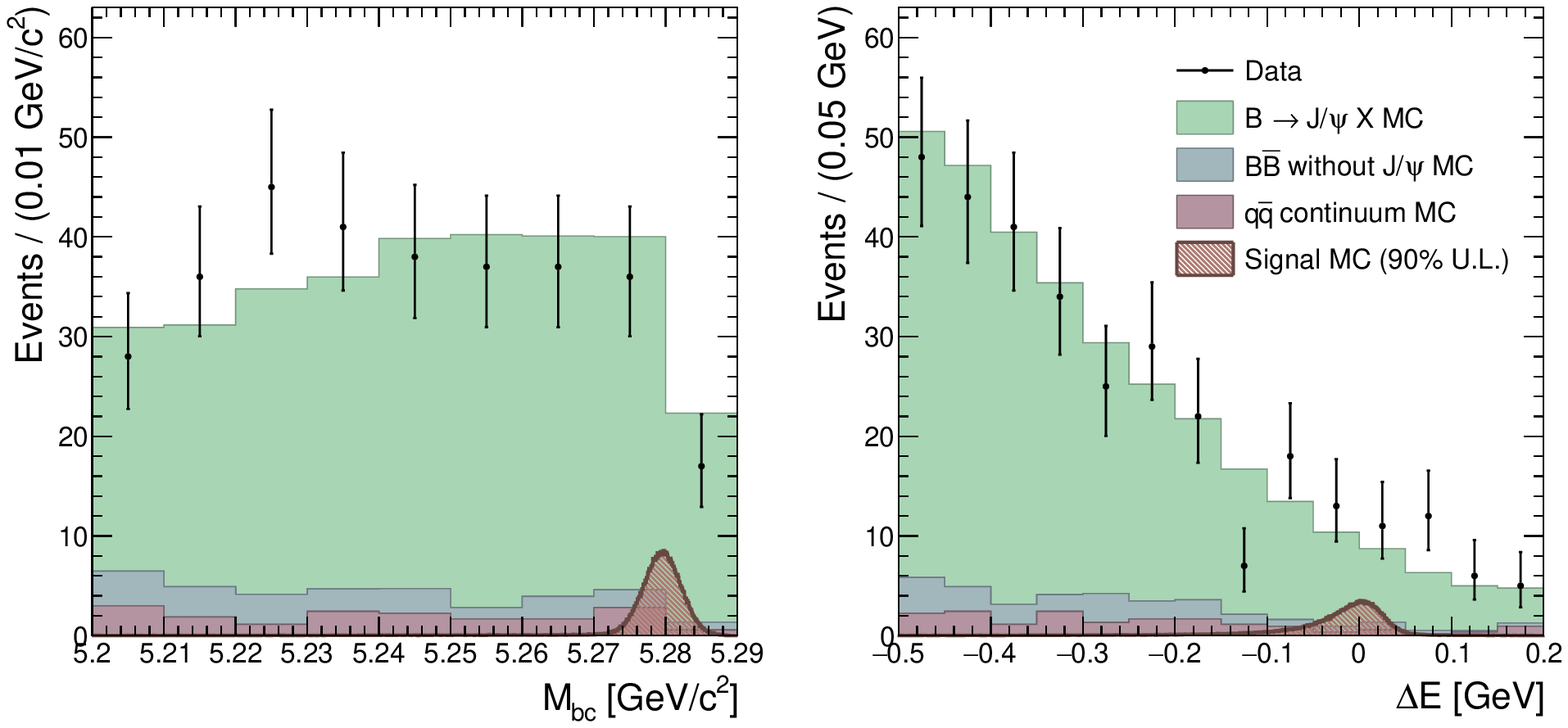}
}
\subfigure[Di\-electron channel.]{
\includegraphics[width=0.5\textwidth]{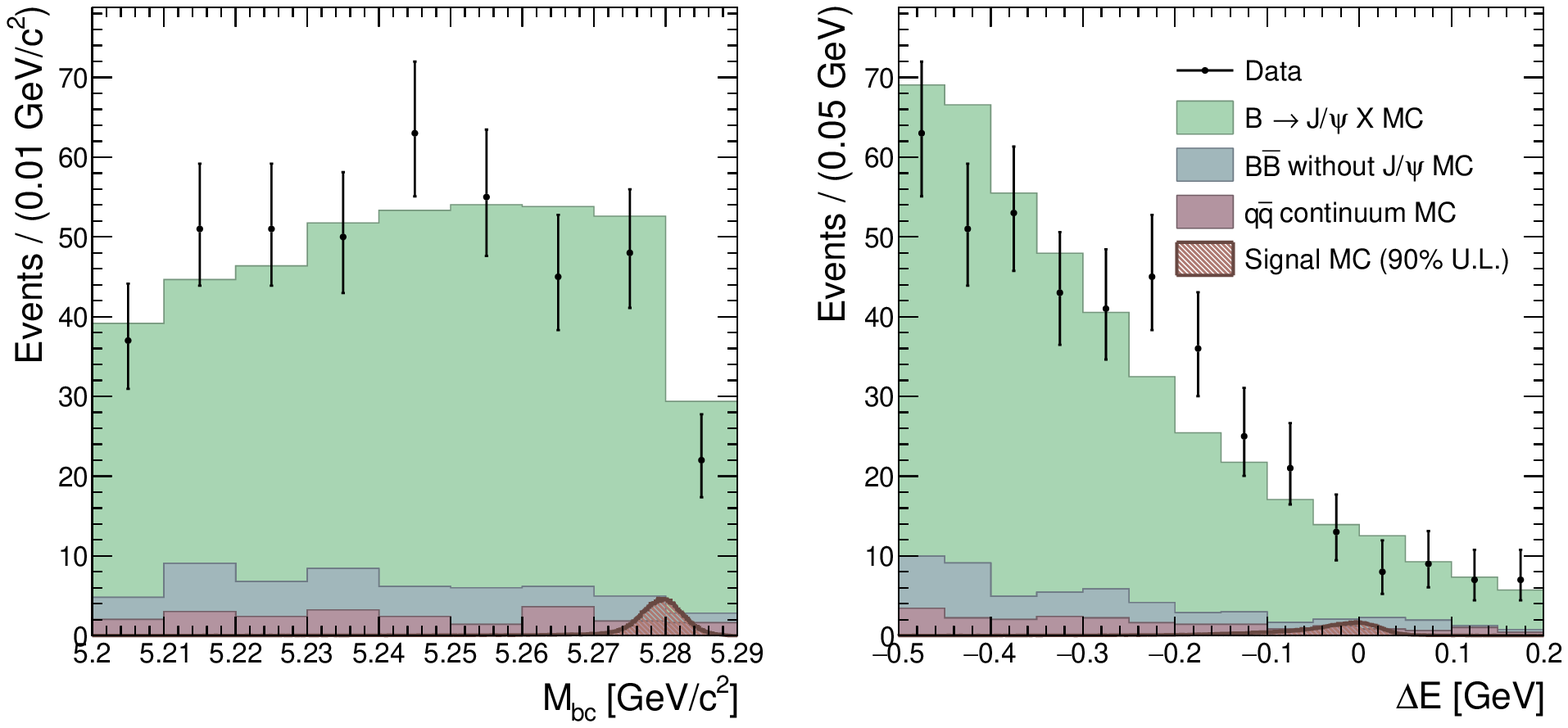}
}
\caption{$M_\text{bc}$ (left) and $\Delta E$ (right) distributions of the selected $B^0 \to X(3872) \gamma$ candidates (data points with error bars), with the estimated background represented as stacked histograms. The components are, from bottom to top: the $q\bar{q}$ continuum background (purple), the $B\bar{B}$ background without $J/\psi$ (blue), and the inclusive $B \to J/\psi X$ background (green).
The signal distribution (hatched brown with thick boundary) is shown corresponding to 90\% C.L. upper limit.}
\label{fig:proj}
\end{figure}

The expected number of background events $N_\text{bkg}$ in the signal region is estimated as
\begin{equation}
N_\text{bkg} = N_\text{sb, data} \times \frac{N_\text{bkg, MC}}{N_\text{sb, MC}},
\end{equation}
where $N_\text{sb, data}$ and $N_\text{sb, MC}$ are the number of data events and the background MC events in the sideband region (selection region with signal region excluded), respectively. 
$N_\text{bkg, MC}$\ is the number of background MC events in the signal region. The ratio of $B \to J/\psi X$ and other backgrounds are fixed to MC expectation.
The expected number of background events in the signal regions are $N_\text{bkg}=9.3$ and $12.1$ for the di\-muon and di\-electron channels, respectively.
The observed number of events in the signal region are $N_\text{evt}=9$ for both di\-muon and di\-electron channels, and the data scatter plots with the signal regions shown as rectangle are shown in Fig.~\ref{fig:scat}.
The projections of the data and the estimated background are displayed in Fig.~\ref{fig:proj}.

\begin{table}[b]
\caption{Summary of results from the counting method.}
\label{table:counting}
\begin{tabular}
{@{\hspace{0.25cm}}l@{\hspace{0.25cm}}   @{\hspace{0.25cm}}c@{\hspace{0.25cm}}  @{\hspace{0.25cm}}c@{\hspace{0.25cm}}  @{\hspace{0.25cm}}c@{\hspace{0.25cm}}}
\hline \hline
Channel & Di\-muon  & Di\-electron & Total\\
\hline
Observed $N_\text{evt}$ & 9 & 9 & 18 \\
Expected $N_\text{bkg}$ & 9.3 &  12.1 & 21.4 \\
90\% U.L.& $9.2\times10^{-7}$ & $6.8\times10^{-7}$ & $5.1\times10^{-7}$ \\
\hline \hline
\end{tabular}
\end{table}

As we find no evidence for the decay $B^0 \to X(3872) \gamma$, we give an upper limit on the branching fraction at 90\% confidence level (C.L.).
We apply the Feldman-Cousins counting method~\cite{FeldmanCousins} using the implementation provided in the \textsc{TRolke} package~\cite{ref:TRolke}, which takes into account separately the uncertainties in the background and the efficiency.
The expected number of background events $N_\text{bkg}$ in the signal region is estimated using the sideband data and the ratio of number of events in the signal region and in the sideband region in the background MC.
The uncertainties in the background levels are studied by comparing the sideband data and the sideband background MC, and are 10.9\% and 17.3\% for the di\-muon and the di\-electron channels, respectively. 
We thus determine the upper limit on the product of the branching fractions $\mathcal{B}(B^0 \to X(3872) \gamma)\times \mathcal{B}(X(3872) \to J/\psi \pi^+ \pi^-)$ to be $5.1\times10^{-7}$ at the 90\% C.L.
The results are summarized in Table \ref{table:counting}.

In conclusion, we have performed a search for the decay $B^0 \to X(3872) \gamma$ based on a data sample of 
$711$ fb$^{-1}$ $e^+ e^-$ collisions
collected by Belle.
No significant signal is found. We set an upper limit on the product of the branching fractions $\mathcal{B}(B^0 \to X(3872) \gamma)\times \mathcal{B}(X(3872) \to J/\psi \pi^+ \pi^-)$ of $5.1\times10^{-7}$ at the 90\% confidence level.

We thank the KEKB group for the excellent operation of the
accelerator; the KEK cryogenics group for the efficient
operation of the solenoid; and the KEK computer group, and the Pacific Northwest National
Laboratory (PNNL) Environmental Molecular Sciences Laboratory (EMSL)
computing group for strong computing support; and the National
Institute of Informatics, and Science Information NETwork 5 (SINET5) for
valuable network support.  We acknowledge support from
the Ministry of Education, Culture, Sports, Science, and
Technology (MEXT) of Japan, the Japan Society for the 
Promotion of Science (JSPS), and the Tau-Lepton Physics 
Research Center of Nagoya University; 
the Australian Research Council including Grants
No. DP180102629, 
No. DP170102389, 
No. DP170102204, 
No. DP150103061, 
No. FT130100303; 
Austrian Science Fund (FWF);
the National Natural Science Foundation of China under Contracts
No.~11435013,  
No.~11475187,  
No.~11521505,  
No.~11575017,  
No.~11675166,  
No.~11705209;  
Key Research Program of Frontier Sciences, Chinese Academy of Sciences (CAS), Grant No.~QYZDJ-SSW-SLH011; 
the  CAS Center for Excellence in Particle Physics (CCEPP); 
the Shanghai Pujiang Program under Grant No.~18PJ1401000;  
the Ministry of Education, Youth and Sports of the Czech
Republic under Contract No.~LTT17020;
the Carl Zeiss Foundation, the Deutsche Forschungsgemeinschaft, the
Excellence Cluster Universe, and the VolkswagenStiftung;
the Department of Science and Technology of India; 
the Istituto Nazionale di Fisica Nucleare of Italy; 
National Research Foundation (NRF) of Korea Grants
No.~2015H1A2A1033649, No.~2016R1D1A1B01010135, No.~2016K1A3A7A09005
603, No.~2016R1D1A1B02012900, No.~2018R1A2B3003 643,
No.~2018R1A6A1A06024970, No.~2018R1D1 A1B07047294; Radiation Science Research Institute, Foreign Large-size Research Facility Application Supporting project, the Global Science Experimental Data Hub Center of the Korea Institute of Science and Technology Information and KREONET/GLORIAD;
the Polish Ministry of Science and Higher Education and 
the National Science Center;
Ministry of Science and Higher Education and Russian Science Foundation (MSHE and RSF), Grant No. 18-12-00226;
the Slovenian Research Agency;
Ikerbasque, Basque Foundation for Science, Spain;
the Swiss National Science Foundation; 
the Ministry of Education and the Ministry of Science and Technology of Taiwan;
and the United States Department of Energy and the National Science Foundation.

\end{document}